\PassOptionsToPackage{unicode}{hyperref}
\PassOptionsToPackage{hyphens}{url}
\documentclass[
  manuscript, nonacm]{acmart}
\usepackage{xcolor}
\makeatletter\@ifclassloaded{acmart}{}{\usepackage{amsmath,amssymb}}\makeatother
\usepackage{iftex}
\ifPDFTeX
  \usepackage[T1]{fontenc}
  \usepackage[utf8]{inputenc}
  \usepackage{textcomp} %
\else %
  \usepackage{unicode-math} %
  \defaultfontfeatures{Scale=MatchLowercase}
  \defaultfontfeatures[\rmfamily]{Ligatures=TeX,Scale=1}
\fi
\usepackage{lmodern}
\ifPDFTeX\else
\fi
\IfFileExists{upquote.sty}{\usepackage{upquote}}{}
\IfFileExists{microtype.sty}{%
  \usepackage[]{microtype}
  \UseMicrotypeSet[protrusion]{basicmath} %
}{}
\makeatletter
\@ifundefined{KOMAClassName}{%
  \IfFileExists{parskip.sty}{%
    \usepackage{parskip}
  }{%
    \setlength{\parindent}{0pt}
    \setlength{\parskip}{6pt plus 2pt minus 1pt}}
}{%
  \KOMAoptions{parskip=half}}
\makeatother
\NewDocumentCommand\citeproctext{}{}
\NewDocumentCommand\citeproc{mm}{%
  \hyperlink{cite.#1}{#2}}
\makeatletter
 \def\@biblabel#1{}
\makeatother
\newlength{\cslhangindent}
\newlength{\csllabelwidth}
\newenvironment{CSLReferences}[2] %
 {\begin{list}{}{%
  \setlength{\itemindent}{0pt}
  \setlength{\leftmargin}{0pt}
  \setlength{\parsep}{0pt}
  \ifodd #1
   \setlength{\leftmargin}{\cslhangindent}
   \setlength{\itemindent}{-1\cslhangindent}
  \fi
  \setlength{\itemsep}{#2\baselineskip}}}
 {\end{list}}
\usepackage{calc}

\usepackage{booktabs}
\usepackage{array}
\newcolumntype{P}[1]{>{\raggedright\arraybackslash}p{#1}}
\usepackage{graphicx}
\usepackage{tikz}
\usepackage{placeins}
\usepackage{needspace}
\usetikzlibrary{arrows.meta,positioning}
\usepackage[normalem]{ulem}
\usepackage{newunicodechar}
\newunicodechar{→}{\ensuremath{\rightarrow}}
\newunicodechar{↔}{\ensuremath{\leftrightarrow}}
\newunicodechar{×}{\ensuremath{\times}}
\newunicodechar{−}{\ensuremath{-}}
\newunicodechar{Δ}{\ensuremath{\Delta}}
\newunicodechar{μ}{\ensuremath{\mu}}
\newunicodechar{±}{\ensuremath{\pm}}
\newunicodechar{≥}{\ensuremath{\geq}}
\newunicodechar{≠}{\ensuremath{\neq}}
\newunicodechar{≈}{\ensuremath{\approx}}
\newunicodechar{✓}{\checkmark}
\newunicodechar{·}{\textperiodcentered}
\newunicodechar{…}{\ldots}


\definecolor{asexpected}{HTML}{1F6F43}
\definecolor{against}{HTML}{B0302A}
\newcommand{\gain}[1]{\textcolor{asexpected}{#1}}
\newcommand{\loss}[1]{\textcolor{against}{#1}}
\usepackage{bookmark}
\IfFileExists{xurl.sty}{\usepackage{xurl}}{} %
\hypersetup{
  pdftitle={Affora},
  hidelinks,
  pdfcreator={LaTeX via pandoc}}

\title{Affora}
\usepackage{etoolbox}
\makeatletter
\providecommand{\subtitle}[1]{%
  \apptocmd{\@title}{\par {\large #1 \par}}{}{}
}
\makeatother
\subtitle{A Design System for Agent-Friendly Interfaces}
\author{Jin Gao}
\orcid{0000-0002-1595-6895}
\affiliation{%
  \institution{Independent Researcher}
  \city{San Francisco}
  \state{CA}
  \country{USA}}
\date{}

\DeclareMathDelimiter{(}{\mathopen}{operators}{"28}{largesymbols}{"00}
\DeclareMathDelimiter{)}{\mathclose}{operators}{"29}{largesymbols}{"01}
\DeclareMathDelimiter{[}{\mathopen}{operators}{"5B}{largesymbols}{"02}
\DeclareMathDelimiter{]}{\mathclose}{operators}{"5D}{largesymbols}{"03}
\DeclareMathDelimiter{/}{\mathord}{letters}{"3D}{largesymbols}{"0E}
\begin{document}
\begin{abstract}

Computer-use agents increasingly operate software designed for people, but interfaces often leave actions or task state unclear to machine readers. We present Affora, a design system that supports both readers while preserving visual freedom and familiar human workflows. Three controlled studies examine component implementations, visual variation, and interaction-design principles. Their findings inform guidance from individual components to complete sites, supported by reusable implementations and executable checks. Agent performance depends on the interaction meaning available through its interface representation; substantial visual variation remains possible when that meaning is preserved. Evaluation on independently authored interfaces shows gains where Affora addresses existing deficits, but limited effects where those deficits are absent or outside its coverage. A workflow case provides preliminary evidence of reduced interaction cost. Affora connects user experience and agent experience through a shared interface rather than a separate agent-only surface.

\end{abstract}
\begin{teaserfigure}
\centering
\includegraphics[width=\textwidth]{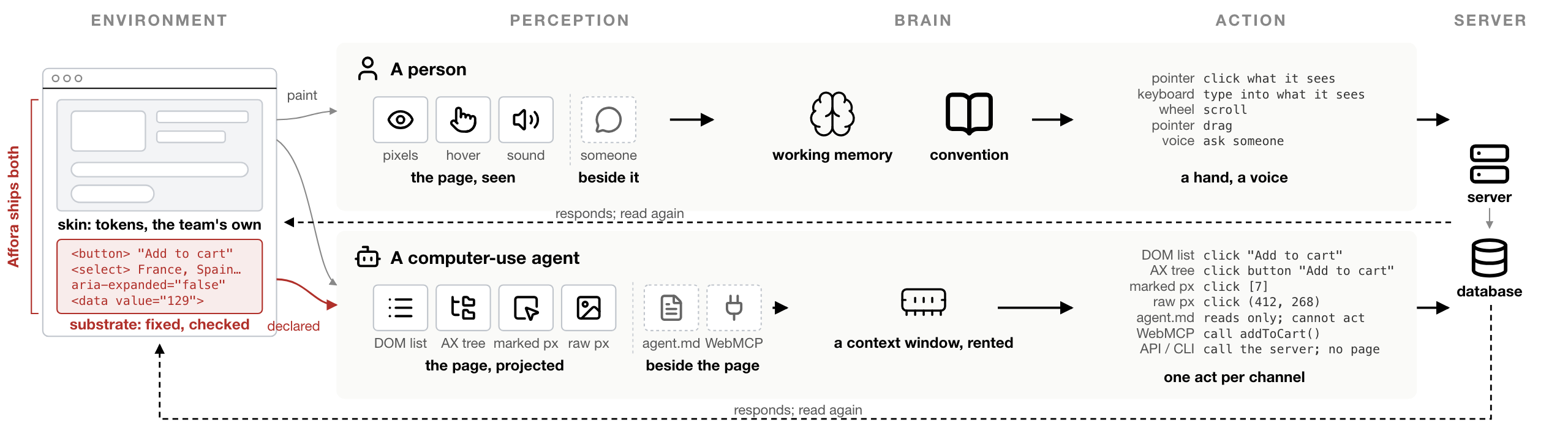}
\Description{A left-to-right diagram comparing how a person and a computer-use agent perceive, reason about, and act on the same interface. The interface contains a visual skin and an underlying declared substrate. The person primarily perceives the rendered interface, while the agent may receive a DOM representation, accessibility tree, marked screenshot, or raw screenshot. Additional agent-only channels such as agent.md and WebMCP bypass parts of the shared interface.}
\caption{\textbf{One interface, two kinds of reader.} People and agents may act on the same application while receiving different representations of its controls, state, and possible actions.}
\label{fig:readers}
\end{teaserfigure}
\maketitle

\subsection{1. Introduction}\label{introduction}

Computer-use agents increasingly operate software interfaces used for everyday tasks. Their performance depends not only on model capability but also on how an interface represents its available actions and current state. A control may appear obvious to a person yet be absent or ambiguous in the representation an agent receives. Improving agent reliability therefore raises a design question: how should the interface itself support an agent reader while remaining familiar to people?

We present \textbf{Affora}, a design system for a shared human--agent interface. Its name draws on J.\ J.\ Gibson's concept of \emph{affordance}, understood as possibilities for action in relation to an actor's capabilities \citeproc{ref-gibson1977affordances}{(Gibson 1977)}. Affora specifies how interaction meaning should remain recoverable from the interface while allowing visual expression to vary. In the web implementation studied here, this structure is carried by the rendered DOM and exposed through the agent's observation channel. The system is intended to fit existing design and development workflows while retaining familiar interfaces for human users.

We investigate this approach through three studies of component implementations, visual variation, and established interaction-design principles. Their findings inform design guidance across component, layout, flow, and site scales, implemented through reusable components and executable checks. We then evaluate transfer to independently authored interfaces and examine interaction cost in a complete workflow. The contribution is a design system together with evidence about where its rules help, where they have little effect, and where their coverage remains incomplete. Affora connects user experience (UX) and agent experience (AX) within one interface; its benefits for human users remain a subject for future evaluation.

\subsection{2. Related work}\label{related-work}

\textbf{Design systems, accessibility, and semantic interfaces.}
Design systems distribute tokens, components, and interaction patterns to preserve consistency and accessibility across products (Putnam, Rose, and MacDonald 2023; Lamine and Cheng 2022). Accessibility standards such as WCAG and ARIA likewise specify how controls, names, relationships, and states should be exposed beyond their visual presentation. These traditions establish that an interface has structure beneath its styling, but they optimize primarily for human use, including use through assistive technologies. Affora builds on that semantic foundation while asking a different question: whether the rendered interface exposes enough persistent, enumerable, and operable structure for an autonomous reader to complete a task. Agents and disabled users are not equated; rather, both make the consequences of an underspecified semantic layer visible (Reid 2024).

\textbf{Computer-use agents and interface grounding.}
Research on computer-use agents has largely asked how an agent can understand and act on interfaces as they are given. Mind2Web, WebArena, VisualWebArena, SeeAct, WebVoyager, BrowserGym, and related systems study grounding through DOM representations, accessibility trees, screenshots, or combinations of these representations (Deng et al. 2023; Zhou et al. 2024; Koh et al. 2024; Zheng et al. 2024; He et al. 2024; Le Sellier De Chezelles et al. 2024). Set-of-Mark prompting similarly alters the agent's visual observation to make targets easier to reference (Yang et al. 2023). Affora reverses the direction of intervention: rather than adapting the reader to an existing interface, it asks which properties of the interface itself should be invariant so that multiple kinds of reader can operate it reliably.

\textbf{Agent-facing representations and action layers.}
Agent-facing documentation and structured tools provide additional ways for agents to understand or operate software. Files such as \texttt{llms.txt} and project instruction files provide information outside the rendered interface \citeproc{ref-howard2024llmstxt}{(Howard 2024)}, while WebMCP exposes callable operations from web applications \citeproc{ref-webmcp2026}{(Walderman et al.\ 2026)}. These mechanisms differ in what they make available to the agent and need not replace the graphical interface. Affora focuses on the structure exposed by the interface itself. Section~6.2 compares concrete implementations on targeted interface failures; Table~\ref{tab:alternatives} describes those interventions rather than guarantees of the underlying standards.

\begin{table}[htbp]
\caption{\textbf{Representations and operations exposed by the evaluated interventions.} Entries describe the implementations used in our comparison, not universal properties of accessibility standards, instruction files, or tool protocols.}
\label{tab:alternatives}
\small
\begin{tabular}{@{}P{0.27\linewidth}P{0.38\linewidth}P{0.29\linewidth}@{}}
\toprule
Intervention & What the agent receives & Where actions execute \\
\midrule
ARIA / structured data & Additional interface metadata & Existing interface \\
Instruction file & Separate textual guidance & Existing interface \\
WebMCP-style tools & Tool descriptions and callable operations & Tool interface \\
Affora & Revised shared-interface structure & Revised interface \\
\bottomrule
\end{tabular}
\end{table}

\textbf{Agent experience and agent-compatible interface design.}
Recent work has begun to treat computer-use agents as a distinct class of interface reader. Goldenberg and Goldenberg discuss \emph{agent experience} by reconsidering Nielsen's usability heuristics for GenAI agents \citeproc{ref-goldenberg2025agentexperience}{(Goldenberg and Goldenberg 2025)}. Rongon et al.\ propose a framework for agent-readable interfaces that emphasizes observable state and explicit interaction meaning \citeproc{ref-rongon2026agentreadable}{(Rongon, Hasan, and Prangon 2026)}. Liu et al.\ experimentally evaluate augmentations of Nielsen's heuristics, finding improved agent task completion and no observable usability regressions in their human evaluation \citeproc{ref-liu2026augmenting}{(Liu et al.\ 2026)}. Together, these works establish interface design itself as a variable in reliable computer use.

Affora extends this direction by expressing agent compatibility through reusable implementations and checks within a design system. It tests principles drawn from multiple interaction-design traditions under readers with different perceptual, memory, and action costs, and applies the resulting guidance across component, layout, flow, and site scales. Section~4.3 examines these differences experimentally.

\subsection{3. Research Questions and Interface Model}\label{research-question-and-methodology}

Affora is motivated by a simple question: \emph{what properties must an interface expose for a computer-use agent to operate it reliably, without constraining how it looks to people?} We address this through three research questions:

\textbf{{RQ1: Where does agent-facing interface failure arise?}}
When computer-use agents fail on ordinary interface components, to what extent is the failure explained by the interface's semantic substrate rather than by model capability or visual presentation?

\textbf{{RQ2: Which visual design decisions can vary without reducing agent performance?}}
If the semantic substrate is held fixed, how freely can designers vary styling and layout while preserving reliable agent interaction?

\textbf{{RQ3: How do established interaction-design principles change when the reader is an agent?}}
Which principles developed for human users still hold, which become more important, and which weaken or reverse when the reader has different perceptual, memory, and action costs?

The three studies in Section~4 address these questions in order. Study~1 compares component implementations to localize the deficit; Study~2 holds the substrate fixed while varying visual design; and Study~3 experimentally re-examines established interaction-design principles for computer-use agents.

\subsubsection{A shared interface, two readers}
\label{shared-interface-two-readers}

A software interface presents more than what it visually renders. We distinguish between what an interface \textbf{paints} and what it \textbf{declares}. The painted layer contains its visual expression---colour, typography, shape, spacing, composition, motion, and other styling decisions. The declared layer contains the controls, names, relationships, state, choices, and task-relevant information represented in a form that can be recovered from the interface itself. We call this declared layer the \textbf{semantic substrate}, or simply the \emph{substrate} below.

People and computer-use agents can operate the same interface while reading these layers differently. A person primarily perceives the rendered interface and interprets its text, spatial organization, visual hierarchy, and learned conventions. An agent instead acts through a projection of the interface, such as a serialised list of interactive elements with roles and names (Zheng et al. 2024; Lù, Kasner, and Reddy 2024), an accessibility tree (Zhou et al. 2024; Le Sellier De Chezelles et al. 2024), a marked screenshot, or raw pixels. These projections expose different portions of the interface and different action spaces. DOM, accessibility-tree, and marked-screen representations explicitly enumerate candidate actions, whereas raw-pixel interaction requires the agent to infer actionable regions visually.

The substrate matters at multiple scales. At component scale, it determines whether an apparent control is represented as an operable control, whether it has a stable name, whether its current state can be recovered, and whether its available choices are exposed. At layout scale, it preserves recoverable relationships among components as their composition changes. At flow scale, it determines whether the interface carries the state and progress needed to continue a task. At site scale, it includes stable mappings between functions, names, and representations across pages. Our use of \emph{substrate} is narrower than the term's use in interaction theory for the computational medium in which information resides (Beaudouin-Lafon, Bødker, and Mackay 2021): here it denotes the machine-readable structure carried by the rendered interface itself.

Figure~\ref{fig:readers} makes this distinction concrete. The same rendered interface may provide rich visual cues to a person while exposing only part of its actionable structure to a machine reader. Conversely, an agent-facing channel such as \texttt{agent.md} or WebMCP may provide information or actions that do not belong to the shared rendered interface at all. Affora focuses on the shared surface: the structure that remains part of the interface both readers use.

\begin{figure}[htbp]
\centering
\includegraphics[width=\linewidth,trim=0 252bp 0 0,clip]{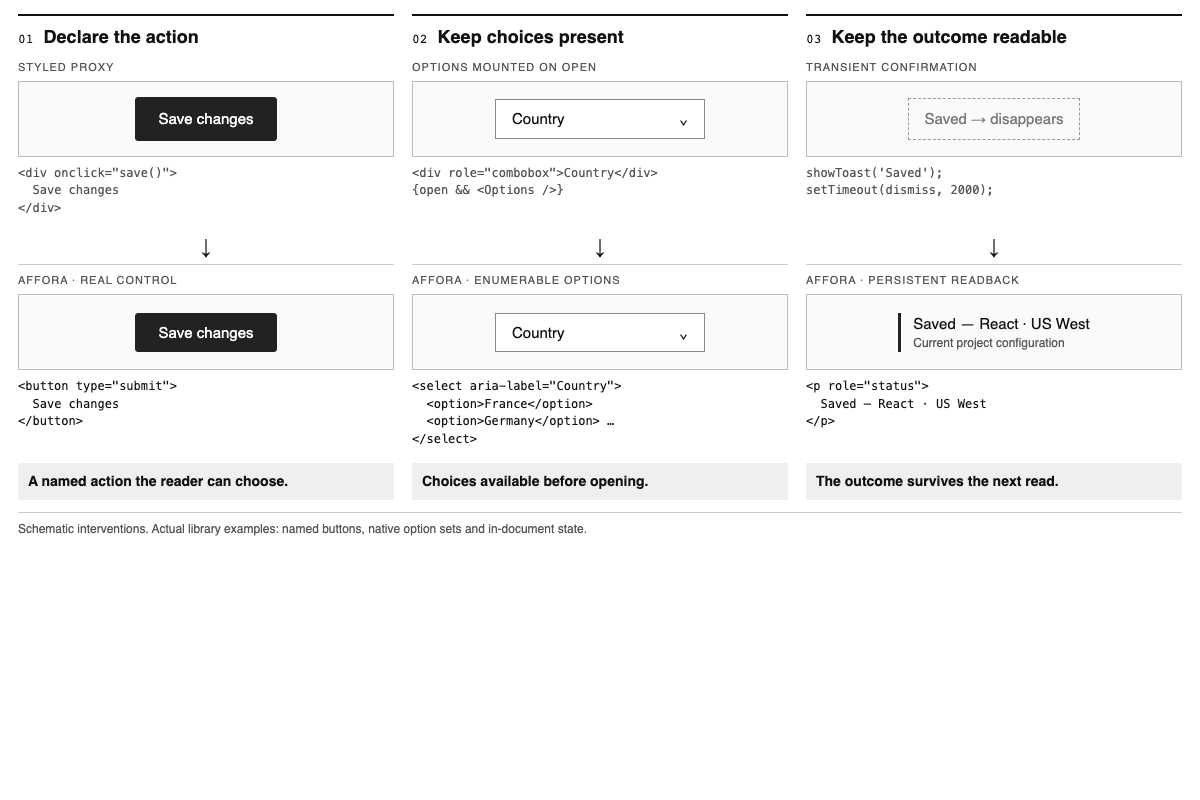}
\Description{Three before-and-after interface interventions: replacing a styled proxy with a named operable control, keeping a set of choices available to the machine-readable representation, and retaining an outcome as persistent state rather than transient feedback.}
\caption{\textbf{Examples of changes to the semantic substrate.} The visual interface can remain similar while its controls, choices, and state become more explicitly represented for machine readers.}
\label{fig:substrate}
\end{figure}

Figure~\ref{fig:substrate} illustrates three representative substrate changes: making an apparent control directly operable, keeping choices available to the machine-readable representation, and retaining an outcome as persistent state. These interventions can alter what an agent can perceive and act upon without requiring a corresponding change in the interface's visual identity.

The substrate is therefore not an agent-only representation. It remains part of the same interface rendered for a person, which creates substantial overlap with accessibility practice: both depend on controls, names, relationships, and state being represented beyond visual styling. Affora builds on that foundation but focuses specifically on whether the shared interface exposes the information a computer-use agent needs to identify available actions, recover task state, and continue interaction through the representations it receives.

In Norman's terms (Norman 2008, 2013), visual appearance can provide a strong signifier to a person while providing little or no corresponding signifier in the representation available to an agent. Affora therefore treats agent-facing interface design as a problem of making action possibilities explicit across machine-readable representations while leaving the visual expression of the interface free to vary.

\subsection{4. What Makes an Interface Agent-Friendly}\label{what-makes-an-interface-agent-friendly}

We examine agent-friendliness as a property of the interface, rather than of agent capability alone. Three studies progressively isolate where it comes from, what can vary without disrupting it, and how it changes established interaction-design assumptions.

\subsubsection{4.1 Study 1: Where interface failures arise}
\label{study-1-where-interface-failures-arise}

\begin{figure}[htbp]
\centering
\includegraphics[width=\linewidth]{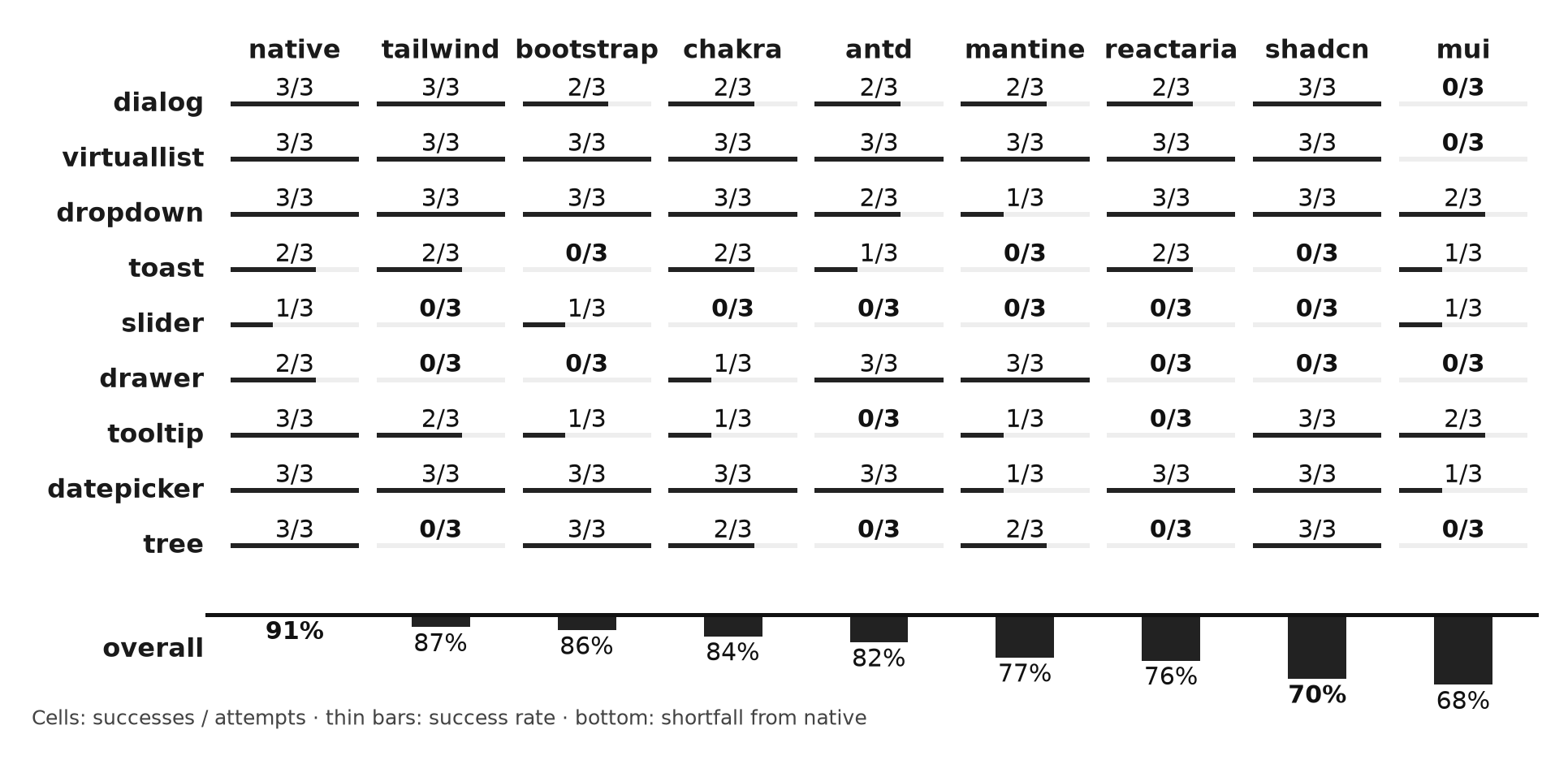}
\Description{A matrix with nine component kinds as rows and nine implementations as columns, ranked by overall agent task success. Each cell reports successes out of three attempts. Different libraries show different failure patterns across component kinds, while plain semantic HTML provides the strongest overall baseline.}
\caption{\textbf{Agent success across component libraries and component kinds.} The same interaction tasks are implemented using native HTML and eight component libraries, revealing systematic differences in agent task completion despite equivalent task content. The matrix shows nine representative component kinds; aggregate results cover the full component set.}
\label{fig:libs}
\end{figure}

\textbf{Design.} We compare implementations of sixty interactive components using native HTML and eight commonly used component libraries, holding task content and goals constant. The libraries represent modern UI abstractions rather than systems designed for agents. We examine how each implementation exposes actionable structure after rendering; Figure~\ref{fig:libs} shows nine representative component kinds.

\textbf{Main result.} Agent performance varies systematically across implementations. Plain semantic HTML reaches 91\%, while the eight component libraries range from 86.7\% to 68.3\% on the mid-tier model (Fig.~\ref{fig:libs}). Failures cluster around recurring representational differences: choices that appear only after interaction, controls whose role or state is difficult to recover, and interaction outcomes that are weakly represented in the rendered structure. In a controlled repair sequence, correcting semantics raises success from 43\% to 67\%, and additionally exposing agent-relevant choices and state raises it to 90\%.

\textbf{Meaning.} The result is not that modern component libraries are poorly designed, but that abstractions optimized for human-facing interfaces do not necessarily preserve the same information for a machine reader after rendering. Two implementations can present equivalent tasks to a person while exposing different DOM structure, state, or action possibilities to an agent. Semantic correctness therefore matters, but is not sufficient: reliable agent interaction also depends on whether relevant controls, choices, and outcomes remain explicit and recoverable in the interface substrate.

\subsubsection{4.2 Study 2: What visual design matters to agents}
\label{study-2-what-visual-design-matters}

\begin{figure}[htbp]
\centering
\includegraphics[width=\textwidth]{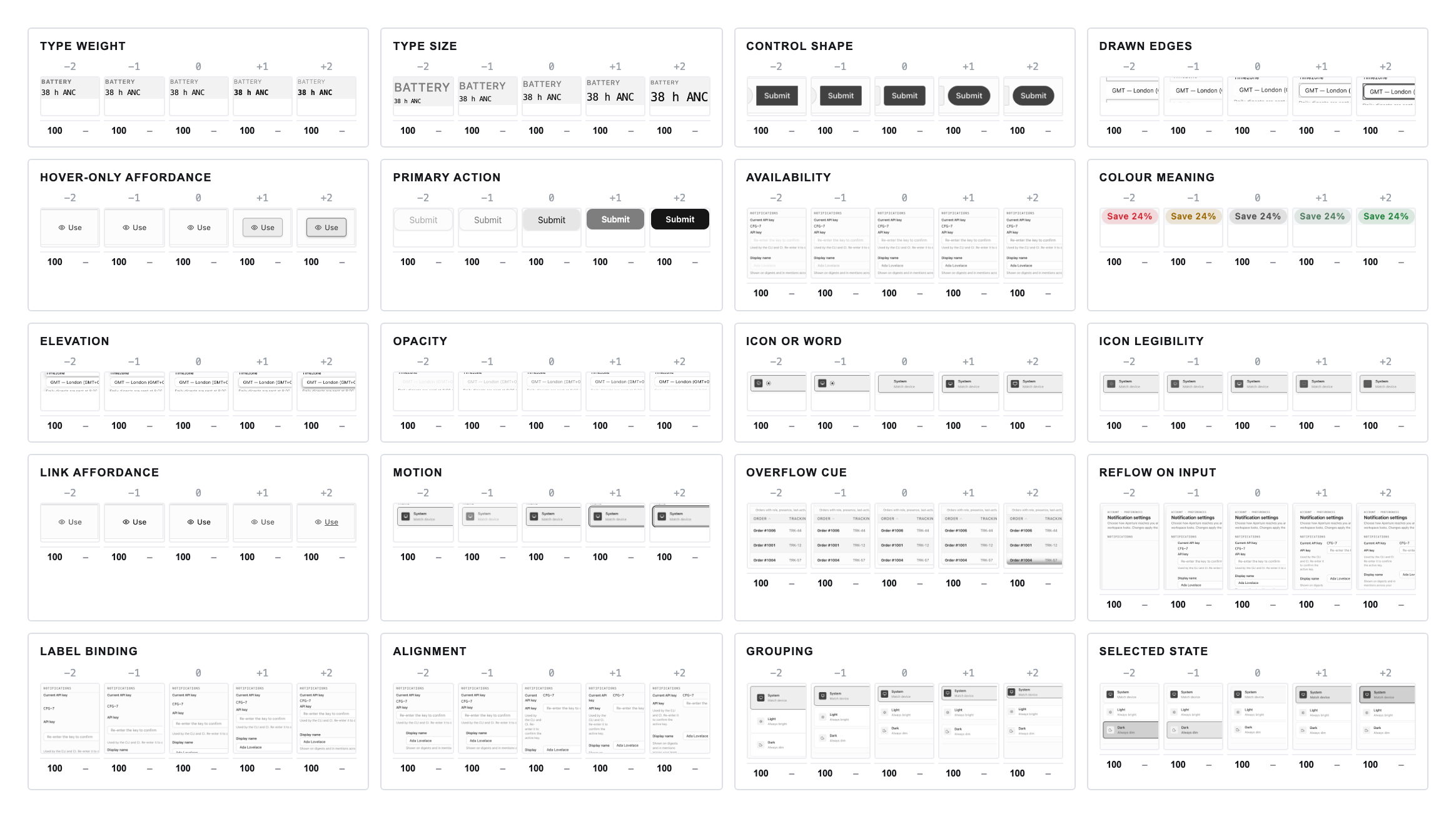}
\Description{A matrix of visual properties at five signed strengths, with descriptive success rates for available runs; missing results are marked by a dash. Treatments range from visual treatments that contradict familiar interaction conventions, through neutral treatments, to treatments that strongly reinforce those conventions. The semantic substrate and task content remain fixed across conditions.}
\caption{\textbf{Visual properties from counter-signaling to reinforcing familiar affordances.} Each family varies one visual property while holding the semantic substrate and task content constant, providing exploratory comparisons of visual signals. Missing model results are not treated as failures.}
\label{fig:signed}
\end{figure}

\textbf{Design.} We study visual design at two levels while holding task content and semantic substrate constant. First, five components are rendered across sixteen themes and six layout archetypes. Second, we vary individual visual properties on a signed scale from treatments that contradict familiar conventions, through neutral treatments, to treatments that reinforce them (Fig.~\ref{fig:signed}). For this exploratory visual-cue analysis, we report complete five-condition blocks for the available Luna marked-pixel run; each block shares a component, property family, and repetition.

\textbf{Main result.} Across the tested themes, DOM-channel completion remains at 99.5\%, matching the native presentation; pixel performance is 92\%, compared with 94\% for native controls. All episodes succeed in the tested extreme-theme and layout combinations in both channels. The exploratory visual-cue sweep is also close to ceiling: completion ranges from 98.7\% to 99.8\% across the five signed strengths, without a monotonic improvement as cues become more reinforcing. These results do not establish a general benefit from stronger visual cues.

\textbf{Meaning.} Within the tested components and tasks, substantial changes in visual identity preserve agent performance when the semantic substrate remains intact. The visual-cue sweep does not justify ranking aesthetics by agent-friendliness. Its marked-pixel reader also receives enumerated targets, so the result should not be generalized to agents that locate actions from raw pixels alone. Visual freedom is therefore supported within the tested observation models, rather than established for every kind of agent reader.

\subsubsection{4.3 Study 3: How canonical interaction-design principles change for computer-use agents}
\label{study-3-the-interaction-design-canon-under-this-reader}

\begin{figure}[htbp]
\centering
\includegraphics[width=\linewidth]{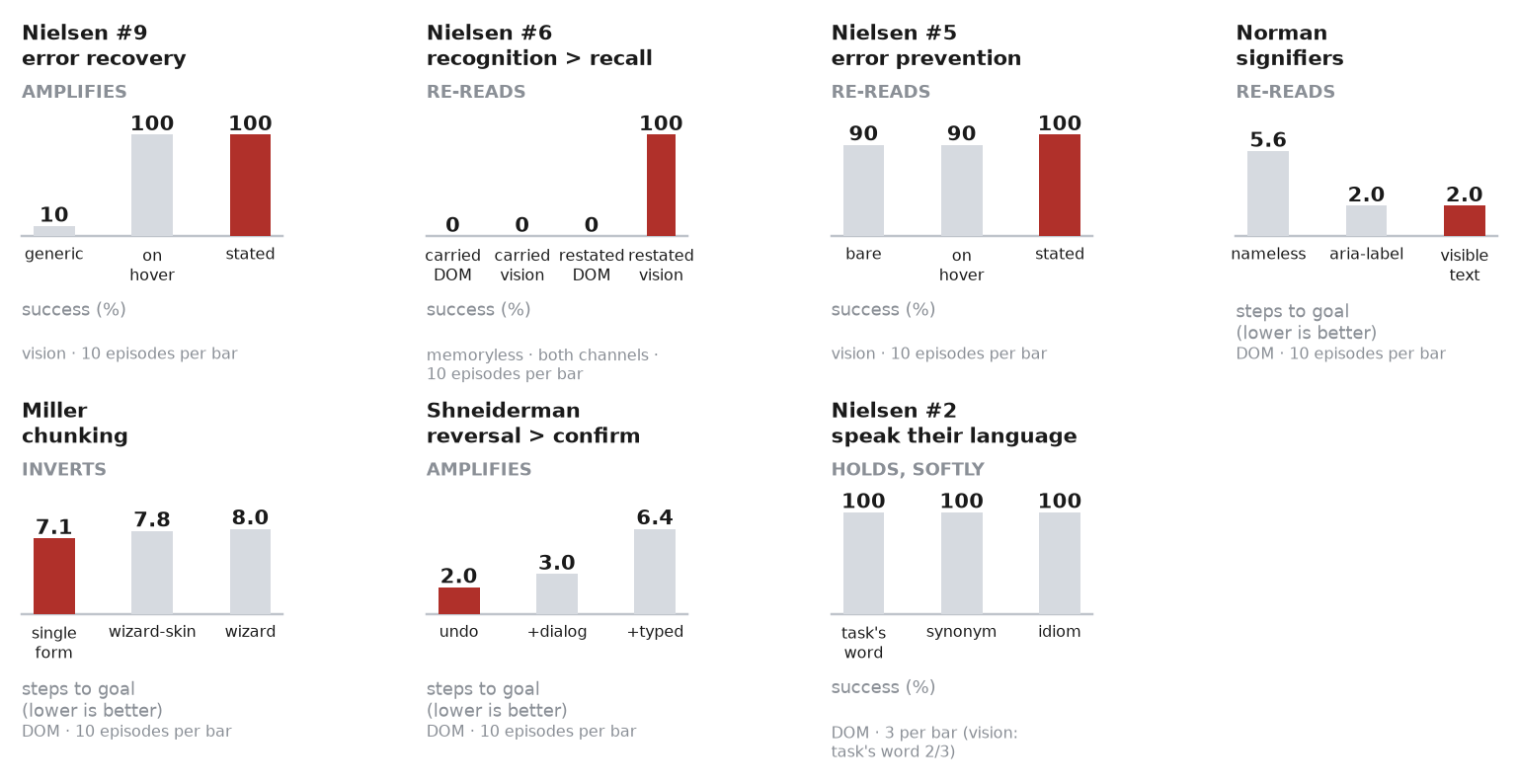}
\Description{Seven small bar charts illustrating Luna measurements for canonical interaction-design principles. Panels report task success or steps to completion, with their observation channel and sample size indicated. The vocabulary panel uses the earlier pilot.}
\caption{\textbf{Illustrative measurements for seven interaction-design principles.} Panels show Luna conditions with the channel and sample size indicated, including an earlier vocabulary pilot; three-model replication results are not shown in this figure.}
\label{fig:canon}
\end{figure}

\textbf{Design.} We translate seven established interaction-design principles into controlled interface comparisons, drawing on Miller \citeproc{ref-miller1956magical}{(1956)}, Nielsen \citeproc{ref-nielsen1994heuristic}{(1994)}, Norman \citeproc{ref-norman2013doet}{(2013)}, and Shneiderman et al.\ \citeproc{ref-shneiderman2016dtui}{(2016)}, holding task information constant while varying the design decision implied by each principle. We measure task completion or interaction cost across three models and two observation channels, targeting ten repetitions per replication condition. Figure~\ref{fig:canon} illustrates conditions from the Luna runs, including the earlier vocabulary pilot. We interpret each principle in relation to the agent's observation, memory, and action model.

\textbf{Main result.} The interaction-design canon does not transfer uniformly to computer-use agents. Explicit remedies improve recovery, while added confirmation increases interaction cost in the tested tasks. Recognition and visual signifiers depend on what the observation channel exposes. Splitting a form into steps adds interaction cost without a demonstrated completion benefit, and explaining disabled states shows no consistent gain across replications. These findings concern observed task behavior, not a direct measurement of cognitive load or the safety value of confirmation.

\textbf{Meaning.} The change of reader changes the mechanism behind familiar design advice. Human-facing principles often assume persistent memory, learned conventions, perceptual continuity, and meaningful motor cost. Computer-use agents may instead re-observe the interface at each step, lose task state across context boundaries, or receive candidate actions already enumerated by their observation channel. Consistency therefore matters when it preserves a mapping the agent must reuse, but less when every screen is interpreted afresh. Recognition matters when the required information is present in the current observation, not merely because it appeared earlier. Confirmation can add unnecessary interaction cost, while explicit recovery becomes more valuable when the agent cannot infer a remedy from context. The implication is not to discard the existing canon, but to restate its principles in terms of the perceptual, memory, and action model of the reader.

\subsection{5. An agent-friendly design
system}\label{an-agent-friendly-design-system}

From the studies, we derive a design methodology for agent-friendly interfaces: make action-relevant structure explicit and machine-readable, preserve visual expression as an independent design layer, and carry task state and interaction meaning beyond individual components. Affora operationalizes this methodology as a design system spanning component, layout, flow, and site scales. Across these scales, Affora follows a common set of design rules:

\begin{itemize}
\setlength{\itemsep}{2pt}
\setlength{\parsep}{0pt}
\setlength{\parskip}{0pt}
\item \textbf{Make task-relevant structure explicit and persistent.} Controls, choices, state, and outcomes should remain available to the reader rather than existing only transiently or in interaction history.

\item \textbf{Align machine-readable and visible structure.} What the interface exposes in the DOM or other machine-readable representations should correspond to what is visibly presented and operated.

\item \textbf{Prefer presence over hidden disclosure.} Information needed for planning should remain discoverable without first requiring hover, expansion, animation, or other transient interaction.

\item \textbf{Preserve semantic continuity.} Recurring functions, task state, and interaction meaning should remain identifiable as the reader moves across components, steps, and pages.

\item \textbf{Keep visual expression flexible.} Colour, typography, shape, composition, and other stylistic decisions may vary as long as they do not remove or contradict the interaction meaning above.

\end{itemize}

Figure~\ref{fig:system} summarizes how these rules are applied throughout the system.

\subsubsection{5.1 Design scales}
\label{design-scales}

\textbf{Component scale.} Affora follows \emph{substrate invariant, skin variable}: controls, names, state, choices, and outcomes remain explicit and operable, while colour, typography, shape, motion, and other stylistic properties may vary. Principles P1--P10 formalize these requirements.

\textbf{Layout scale.} Layout concerns how components are composed while preserving recoverable relationships and access to task-relevant content. It applies existing requirements, particularly P5 on semantic hierarchy and P9 on structural correspondence, across a page rather than introducing a separate family of principles. Layout transformations are assessed through page-level preservation checks and the theme--layout comparisons in Study~2. The evidence status of P5 and P9 remains observational; the layout experiment does not validate every possible composition.

\textbf{Flow scale.} The interface must carry the state a reader needs to continue a task, including progress, prior choices, requirements, recovery paths, and completion. Designers remain free to vary pacing, screen count, and presentation order. Principles F1--F7 and reusable flow patterns encode these requirements.

\textbf{Site scale.} Affora preserves mappings that a reader may need to reuse across pages, such as stable names and representations for recurring functions. It does not require visual uniformity across the product; the constraint is semantic continuity. Principles SC1--SC3 define this scope.

\subsubsection{5.2 System artifacts}
\label{system-artifacts}

Affora is delivered as both a set of plug-and-play UI building blocks and a set of software design guidelines for agent-friendly interfaces. Teams can adopt the reference components directly or reproduce the same rules in their existing stack. The components separate semantic structure from visual expression, allowing Affora to adapt to different visual themes as long as the underlying design constraints are preserved.

\textbf{Reusable UI components.} Affora provides reference implementations for common interface elements, including forms, selection controls, dialogs, tables, and product-oriented components. Each embeds the required semantic substrate by default, so teams do not need to reconstruct agent-facing behavior for every interface.

\textbf{Design guidance and theming.} The library is paired with component-, layout-, flow-, and site-level design rules. Style and layout tokens allow colour, typography, shape, density, and composition to vary while keeping task-relevant controls, state, and relationships intact. Affora therefore defines constraints on interaction meaning rather than prescribing a single visual language.

\textbf{AI-assisted implementation.} Because the design rules are explicit and the reference components are distributed as source, they can be used directly in AI-assisted coding workflows. A coding agent can reuse established components and patterns rather than repeatedly inferring the intended semantics of custom interface code. By making controls, state, choices, and outcomes explicit by construction, Affora also removes a class of interface-induced ambiguities that would otherwise surface later as agent failures and debugging work.

\begin{figure}[htbp]
\centering
\includegraphics[width=\linewidth,height=0.87\textheight,keepaspectratio]{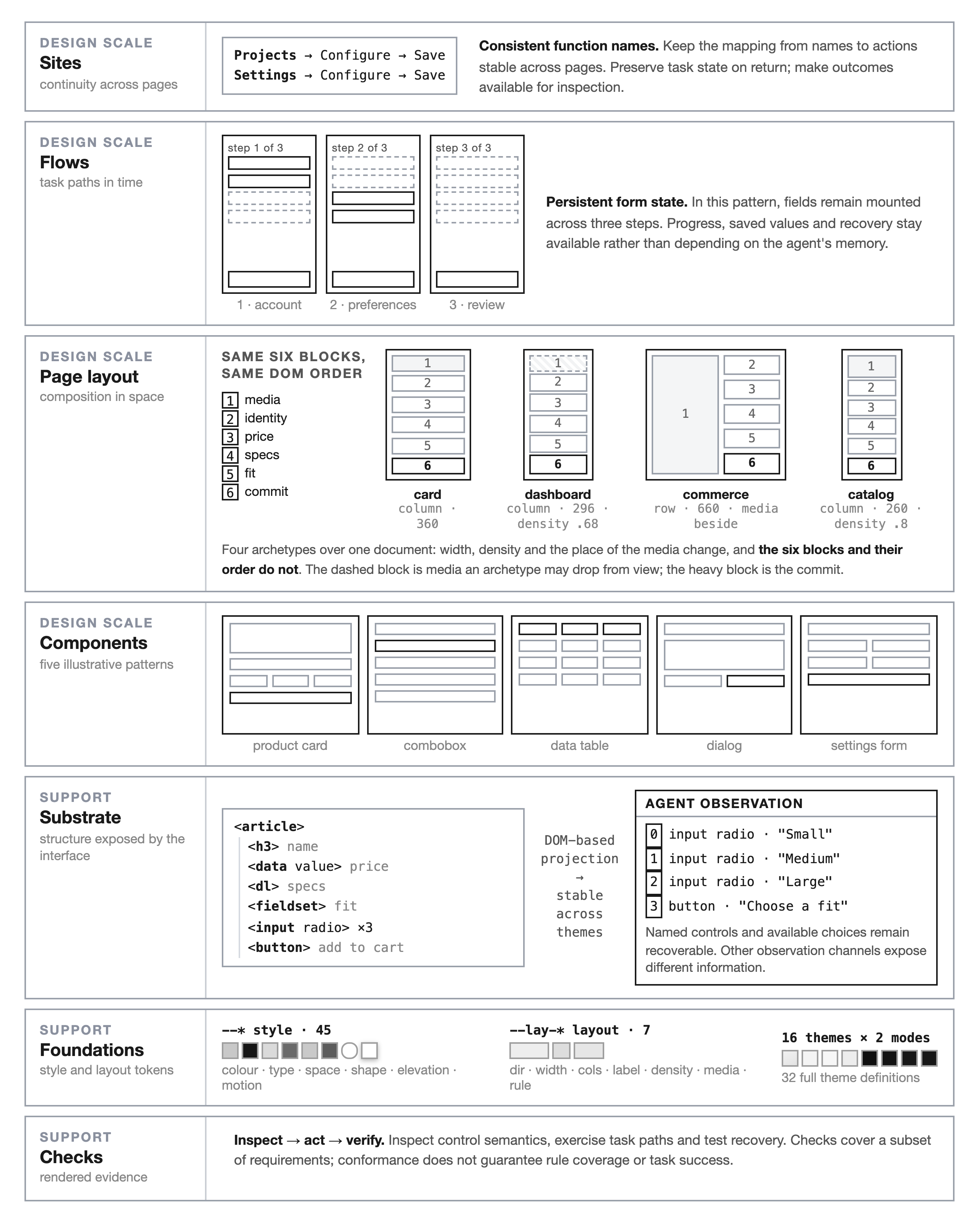}
\Description{The complete Affora system illustrated through site navigation, a three-step form, four page layouts, five example components, a DOM-to-agent representation, visual tokens, and executable checks. These retain the original system diagram's wireframes.}
\caption{\textbf{Affora across four design scales.} Component, layout, flow, and site guidance specifies how interaction meaning is preserved as interfaces are composed. The semantic substrate carries machine-readable structure, tokens support visual variation, and executable checks cover a subset of the design requirements.}
\label{fig:system}
\end{figure}

Figure~\ref{fig:system} relates the design scales to the system's supporting mechanisms.

\begin{figure}[htbp]
\centering
\includegraphics[width=\textwidth]{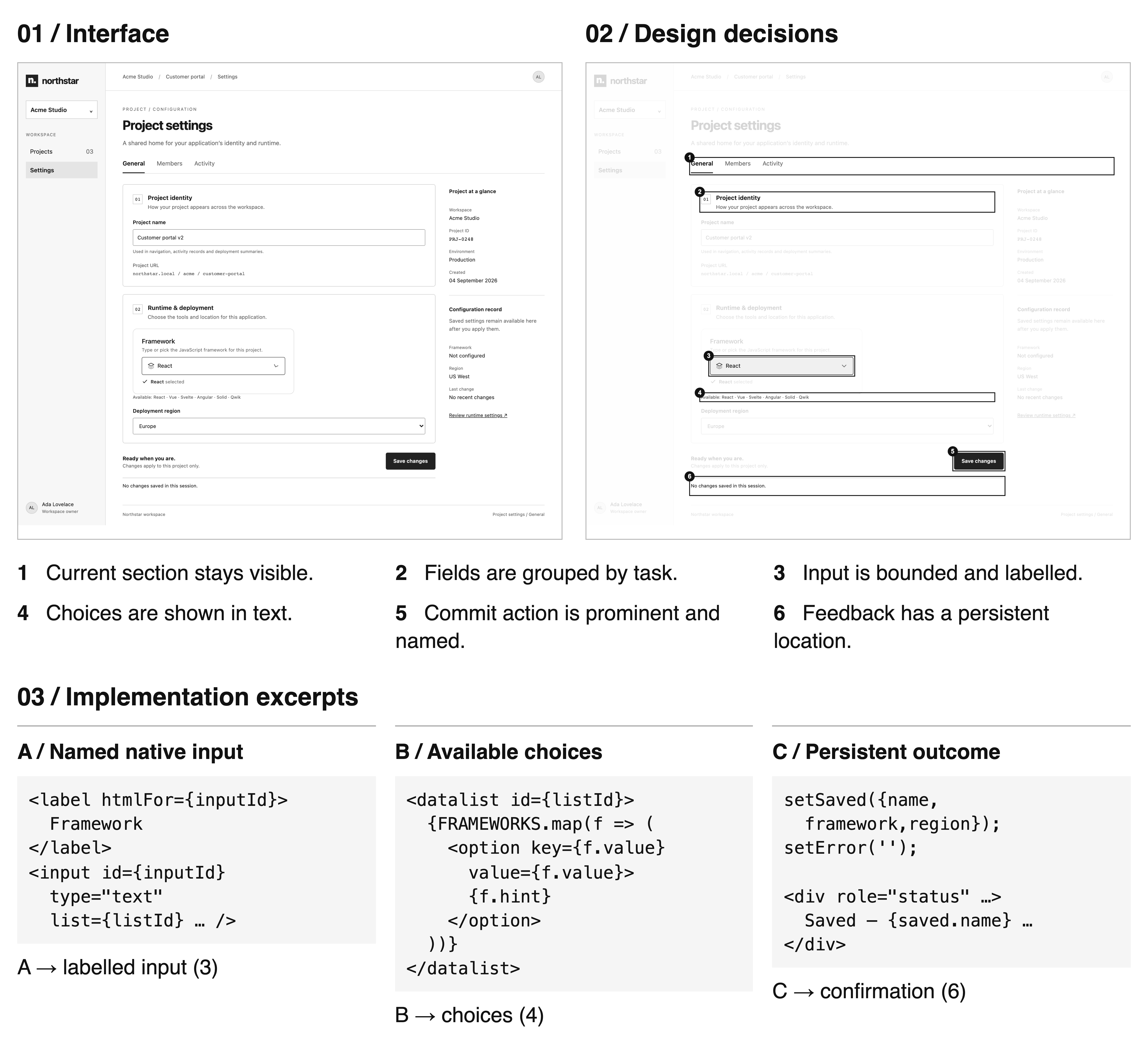}
\Description{The full project-settings page, the same page with six visual annotations, and component/application source excerpts. The annotations locate section context, grouping, input boundaries, visible choices, the save action and feedback. Code shows the native labelled input, datalist option set and saved state.}
\caption{\textbf{Affora in a working application.} The page, its visual affordances and its implementation.}
\label{fig:application}
\end{figure}

Fig.~\ref{fig:application} shows the system in a working application.
Northstar is an author-built project-settings case using the Affora
combobox and theme tokens. The same page appears with visual annotations
and source excerpts; these are implementation examples, not additional
performance measurements.

\subsubsection{5.3 From guidelines to executable constraints}
\label{from-guidelines-to-executable-constraints}

Affora turns part of its design guidance into constraints that software can inspect directly. Component-level checks test whether controls are operable and named, state is recoverable, choices remain enumerable, and machine-readable targets align with the visible interface. Page-level checks additionally verify that transformations preserve information, interaction opportunities, and behavior.

These constraints can also be exposed in machine-readable form to AI-assisted development workflows. Coding agents can consume Affora's rules alongside the component library---for example through project instructions or \texttt{skill.md}-style guidance---and apply them while generating, modifying, or debugging interfaces. This creates a path from post-hoc compliance checking toward enforcing agent-friendly structure during implementation itself.

Passing the checks is a conformance floor rather than a guarantee of task success: it establishes that required interface properties are present, while model capability, visual grounding, flow state, and application behavior may still affect performance. The full mapping from principles to checks and evidence status is given in the appendix, \hyperref[principle-index]{Principle index}.

\subsection{6. Evaluation}
\label{evaluation}

We evaluate whether shared-interface interventions address the same targeted failures as alternative agent-facing mechanisms, whether the rules transfer to independently authored interfaces, and how rule coverage differs from conformance. A separate workflow case examines interaction cost.

\textbf{Execution harness and reproducibility.} All reported episodes run through our authored agent-evaluation harness. It provides a common execution loop for computer use and structured tool use, ReAct-style observe--reason--act cycles, bounded retry after failed actions, and task-scoped memory across steps. We instantiate this harness with the OpenAI ChatGPT 5.6 Luna and ChatGPT 5.6 Terra variants, both at the low reasoning tier, referred to throughout as \emph{Luna} and \emph{Terra}. For every episode, the harness records the model and observation configuration, action and retry trace, terminal application state, and evaluator verdict. Experiments ran locally on macOS~26.2 on a MacBook Pro with an Apple M4 Pro and 48~GB unified memory, using Node.js~v25.2.1, Python~3.14.7, and Docker~29.7.2. We deployed WebArena's Magento storefront and WebShop in local Docker containers, with environment snapshots and task configurations fixed within each comparison.

A \emph{task} or \emph{goal} is a unique specification; an \emph{episode} is one agent attempt under a model and observation condition. Tables report successful episodes over recorded attempts. Results are not pooled across environments, and differing denominators are retained rather than treated as matched pairs.

\subsubsection{6.1 Evaluation setup and cases}
\label{evaluation-setup-and-cases}

Five authored domains support comparison of Affora with ARIA and structured-data augmentation, an \texttt{agent.md} read layer, and a WebMCP-style action layer. WebArena's Magento storefront, WebShop, and four independently authored applications test transfer beyond interfaces built for Affora. The release workflow is an author-built adaptation of a community template and is evaluated separately.

Magento provides a controlled transport test: we introduce a component-level deficit into an otherwise functional application and ask whether the published rule restores the lost behavior. WebShop and the four applications contribute deficits already present in their interfaces. Before agent runs, application tasks are classified by whether the target is visible, predictably disclosed by its label, or hidden behind a label that does not predict its contents.

Within each comparison, both arms use the same recorded-state success criterion. Page-level gates assess whether a rewrite preserves the original information, available interactions, and external effects. These gates establish transformation validity, not a guarantee of agent success.

\subsubsection{6.2 Comparison with alternative interventions}
\label{comparison-with-alternative-interventions}

On the primary comparison set, Affora and the instruction-file condition both reach full completion, an improvement of approximately 23 percentage points over baseline (Table~\ref{tab:interventions}). Affora also reaches full completion on the harder set, while the instruction-file condition has one failure. ARIA and structured-data augmentation produces no consistent improvement in these conditions.

The WebMCP-style condition reaches full completion on its applicable action set. Its different denominator means that this result is not a matched comparison with the other interventions. The findings support the value of making required information or actions available through a representation the agent consumes, without establishing that one mechanism is universally superior.

\begin{table}[htbp]
\centering
\caption{\textbf{Completion under alternative interventions.} Entries are successful episodes / attempts (rounded percentage). The WebMCP-style action set has different coverage and is shown separately.}
\label{tab:interventions}
\small
\begin{tabular*}{\linewidth}{@{\extracolsep{\fill}}P{0.50\linewidth}cc@{}}
\toprule
Intervention & Primary set & Harder set \\
\midrule
Baseline & 46/60 (77\%) & 28/33 (85\%) \\
ARIA and structured data & \loss{45/60 (75\%)} & \gain{29/33 (88\%)} \\
Instruction file (\texttt{agent.md}) & 60/60 (100\%) & 32/33 (97\%) \\
Affora & \gain{60/60 (100\%)} & \gain{33/33 (100\%)} \\
\addlinespace[6pt]
\midrule
WebMCP-style tools (applicable action set) & -- & 55/55 (100\%) \\
\bottomrule
\end{tabular*}
\smallskip
\parbox{\linewidth}{\footnotesize Green marks a favourable change; red an unfavourable change. Black denotes no change or no matched comparison. Colours do not indicate statistical significance.}
\end{table}

\subsubsection{6.3 Transfer and rule coverage}
\label{transfer-to-independent-interfaces}

Across independent interfaces, improvement tracks the presence of the deficit a rule addresses. On Magento, introducing a targeted deficit lowers completion by approximately eight percentage points. Applying the corresponding rewrite restores completion to approximately the original level (Table~\ref{tab:transfer}). Rewriting already-sound components produces no improvement.

WebShop distinguishes rule coverage from faithful implementation. The published rules leave its native hidden-radio failure unchanged. Extending the rule set to cover that idiom raises full-reward completion by approximately 61 percentage points. This extension is a diagnostic response to a newly encountered pattern, not evidence that the original rule set already generalized to it.

In the administrative applications, the largest gains occur for targets behind labels that do not predict their contents. Results for predictably labelled groups are mixed, while already-visible targets show little change. MUI and Atomic CRM are near-null at component level. Table~\ref{tab:transfer-details} reports the application and task-class breakdown, including negative changes. Denominators are recorded attempts and can differ between arms; unmatched attempts are not treated as paired evidence.

\begin{table}[htbp]
\centering
\caption{\textbf{Transfer and coverage in independent environments.} Successful episodes / attempts (rounded percentage). Magento's existing component comparison is separate from the three-model deletion--restoration experiment; WebShop distinguishes the published rules from a subsequent coverage extension.}
\label{tab:transfer}
\small
\begin{tabular*}{\linewidth}{@{\extracolsep{\fill}}P{0.50\linewidth}rr@{}}
\toprule
Comparison & Before & After \\
\midrule
\multicolumn{3}{@{}l}{\textbf{Magento: existing component comparison}} \\
Already-sound components & 7/40 (18\%) & 7/40 (18\%) \\
Untouched control tasks & 3/40 (8\%) & 5/40 (13\%) \\
\addlinespace[7pt]
\multicolumn{3}{@{}l}{\textbf{Magento: three-model deletion--restoration}} \\
Magento: introduce targeted deficit & 33/216 (15\%) & \gain{16/216 (7\%)} \\
Magento: repair targeted deficit & 16/216 (7\%) & \gain{34/216 (16\%)} \\
Unaffected task pairs (287 pairs) & \multicolumn{2}{c}{no systematic movement} \\
\addlinespace[7pt]
\multicolumn{3}{@{}l}{\textbf{WebShop: rule coverage}} \\
WebShop: published rules & 9/117 (8\%) & 9/117 (8\%) \\
WebShop: extend hidden-radio coverage & 9/117 (8\%) & \gain{80/117 (68\%)} \\
\bottomrule
\end{tabular*}
\smallskip
\parbox{\linewidth}{\footnotesize Green marks a favourable change; red an unfavourable change. Black denotes no change or no matched comparison. Colours do not indicate statistical significance. For the deliberate deficit, green marks the expected decrease.}
\end{table}

\begin{table}[htbp]
\centering
\small
\caption{\textbf{Transfer by application and target class.} Successful episodes / attempts (rounded percentage). Three models are represented for shadcn-admin, Ant Design Pro, and Atomic CRM; two for the MUI template.}
\label{tab:transfer-details}
\begin{tabular*}{\linewidth}{@{\extracolsep{\fill}}P{0.50\linewidth}rr@{}}
\toprule
Application / target class & Baseline & Affora \\
\midrule
\multicolumn{3}{@{}l}{\textbf{shadcn-admin: 28 task specifications}} \\
Predictably labelled group & 70/89 (79\%) & \loss{66/89 (74\%)} \\
Label does not predict target & 16/71 (23\%) & \gain{46/72 (64\%)} \\
Already visible & 82/88 (93\%) & \loss{81/90 (90\%)} \\
\addlinespace[7pt]
\multicolumn{3}{@{}l}{\textbf{Ant Design Pro: 21 task specifications}} \\
Predictably labelled group & 0/79 (0\%) & \gain{11/78 (14\%)} \\
Label does not predict target & 12/18 (67\%) & \gain{18/18 (100\%)} \\
Already visible & 86/86 (100\%) & 89/89 (100\%) \\
\addlinespace[7pt]
\multicolumn{3}{@{}l}{\textbf{MUI dashboard template: 20 task specifications}} \\
All target classes & 95/120 (79\%) & \gain{97/120 (81\%)} \\
\addlinespace[7pt]
\multicolumn{3}{@{}l}{\textbf{Atomic CRM: 11 task specifications}} \\
All target classes & 71/98 (72\%) & \loss{70/99 (71\%)} \\
\bottomrule
\end{tabular*}
\smallskip
\parbox{\linewidth}{\footnotesize Green marks a favourable change; red an unfavourable change. Black denotes no change or no matched comparison. Colours do not indicate statistical significance.}
\end{table}

\subsubsection{6.4 Conformance and workflow efficiency}
\label{conformance-and-workflow-efficiency}

\textbf{Conformance.} Preservation gates and rule checks answer different questions. The former test whether the task remains equivalent after a rewrite; the latter inspect whether the interface satisfies specified design requirements. Neither establishes that the requirements cover every interaction pattern or that an agent will complete the task. WebShop illustrates the coverage limit, while near-null component rewrites show why passing component checks cannot establish the quality of an entire workflow. We do not report a separate quantitative validation of the check suite here.

\textbf{Workflow design.} We compare baseline and Affora versions of an author-built release workflow using two models and two observation channels. Correct publication and completion of the final verification are assessed separately. Costs are compared only for paired episodes in which both versions complete the task.

\textbf{Main result.} Affora reduces action counts by approximately 21--40\% and total token use by 24--41\% across the four model--channel conditions (Table~\ref{tab:workflow}). Reasoning-token changes are mixed. Completion improves in the Luna vision condition and remains unchanged in the others. With only one or two successful pairs per condition, the results provide preliminary evidence of lower interaction cost in this workflow, not a general estimate of efficiency gains.

\begin{table}[htbp]
\centering
\caption{\textbf{Release-workflow completion and interaction cost.} Arrows denote baseline $\rightarrow$ Affora. Completion uses all recorded episodes; cost totals use only pairs where both arms succeed. Model--channel conditions are not pooled.}
\label{tab:workflow}
\fontsize{8.5}{10.5}\selectfont
\setlength{\tabcolsep}{1pt}
\renewcommand{\arraystretch}{1.1}
\begin{tabular*}{\linewidth}{@{\extracolsep{\fill}}lcccc@{}}
\toprule
 & \multicolumn{2}{c}{Vision} & \multicolumn{2}{c}{DOM/AX} \\
\cmidrule(lr){2-3}\cmidrule(l){4-5}
Metric & Luna & Terra & Luna & Terra \\
\midrule
Task completion & 1/2 $\rightarrow$ \gain{2/2 (+100\%)} & 2/2 $\rightarrow$ 2/2 (0\%) & 2/2 $\rightarrow$ 2/2 (0\%) & 2/2 $\rightarrow$ 2/2 (0\%) \\
Correct publication & 2/2 $\rightarrow$ 2/2 (0\%) & 2/2 $\rightarrow$ 2/2 (0\%) & 2/2 $\rightarrow$ 2/2 (0\%) & 2/2 $\rightarrow$ 2/2 (0\%) \\
\addlinespace[6pt]
Successful pairs & 1 & 2 & 2 & 2 \\
Actions & 50 $\rightarrow$ \gain{34 (−32\%)} & 98 $\rightarrow$ \gain{77 (−21\%)} & 111 $\rightarrow$ \gain{67 (−40\%)} & 108 $\rightarrow$ \gain{77 (−29\%)} \\
Total tokens (k) & 196.5 $\rightarrow$ \gain{123.7 (−37\%)} & 404.5 $\rightarrow$ \gain{308.0 (−24\%)} & 345.8 $\rightarrow$ \gain{203.2 (−41\%)} & 388.0 $\rightarrow$ \gain{278.1 (−28\%)} \\
Reasoning tokens & 854 $\rightarrow$ \gain{558 (−35\%)} & 608 $\rightarrow$ \loss{925 (+52\%)} & 2,219 $\rightarrow$ \gain{929 (−58\%)} & 602 $\rightarrow$ \loss{620 (+3\%)} \\
\bottomrule
\end{tabular*}

\smallskip
\parbox{\linewidth}{\footnotesize Parentheses show percentage change relative to baseline, rounded to whole percentages; 1/2 to 2/2 is +100\% relative (+50 percentage points). Costs are totals over successful pairs; total tokens include reasoning tokens. Green marks higher completion or lower cost, red the reverse, and black no change. Colours do not indicate statistical significance.}
\end{table}

\begin{figure}[htbp]
\centering
\includegraphics[width=\textwidth,height=0.88\textheight,keepaspectratio]{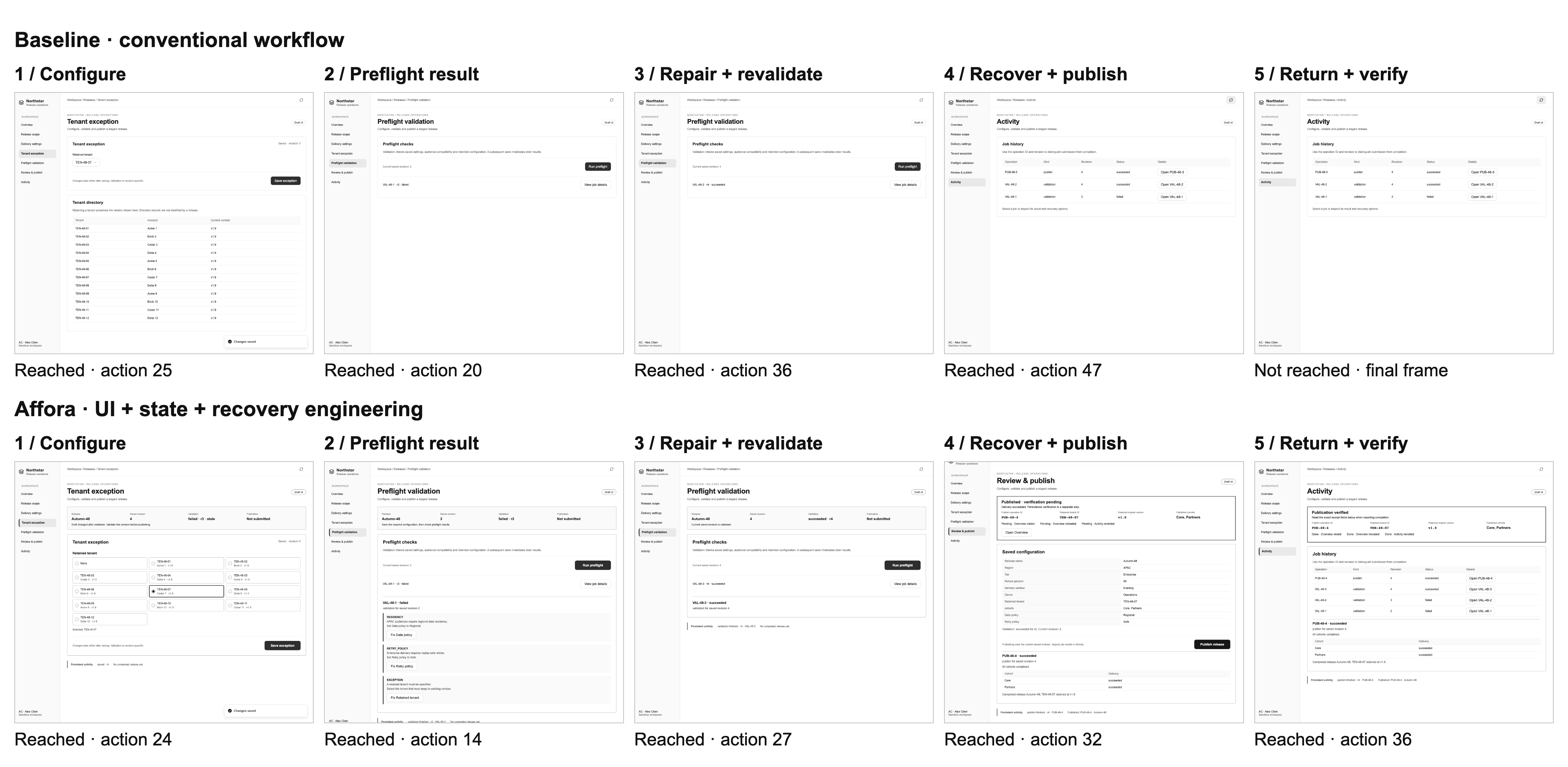}
\Description{Two rows of five full-page observations from fixed Luna vision-agent episodes in release-v5. Baseline is above full Affora. Columns group configuration, preflight result, repair and revalidation, recovered publication, and final verification. Baseline reaches correct publication but not final task completion; Affora reaches all checkpoints. Action numbers reveal the actual order. Model-level aggregate results are reported separately in Table 5.}
\caption{\textbf{From publication to verified completion.} A fixed illustrative pair from the release workflow in Table~\ref{tab:workflow}. Five checkpoint categories expose the UI and state-engineering differences; aggregate evidence belongs to the table.}
\label{fig:showcase}
\end{figure}

\textbf{Visualising the full-workflow case.} Figure~\ref{fig:showcase} shows how the interface changes appear along a fixed baseline--Affora trajectory. Both episodes reach correct publication, but only the Affora episode completes final verification. The figure illustrates the recorded interaction sequence; Table~\ref{tab:workflow} provides the aggregate results.

\FloatBarrier

\subsection{7. Discussion and conclusion}
\label{discussion-and-conclusion}

Affora suggests a broader design position: as agents become software operators, interfaces increasingly serve both human and machine readers. Agent-friendly design should therefore improve machine operation without making software less legible, observable, or controllable for people.

\subsubsection{7.1 Shared and observable agent experience}
\label{shared-observable-agent-experience}

Affora treats agent experience as a property of the software environment, not only of the agent. Explicit controls, persistent state, recoverable outcomes, and visible progress help agents operate an interface, but they can also make autonomous activity easier for people to inspect.

This matters because machine actions may occur faster than people can follow them step by step. Human supervision therefore cannot depend only on reading an execution trace. A shared interface can instead expose the consequential state of the work: what changed, what remains incomplete, what failed, and what can be reversed. We treat this supervisability benefit as a design implication rather than a measured claim, since the present work does not include a human study.

\subsubsection{7.2 Efficiency as an interface property}
\label{efficiency-as-interface-property}

Affora also suggests that agent efficiency is partly an interface property. When state, available actions, progress, and recovery information are explicit, an agent can spend fewer interaction cycles rediscovering context, probing hidden controls, or repeating actions whose effects are unclear. In the end-to-end workflow evaluation, Affora reduces both action count and total token use across all four model--channel conditions, although reasoning-token use is mixed.

This shifts part of the efficiency problem from model optimization to interface design. A more legible interface can reduce unnecessary observations, actions, and context consumption even when the underlying agent is unchanged. In deployed systems, this may translate into lower inference cost and latency, although wall-clock savings were not directly measured in our experiments.

\subsubsection{7.3 Shared interfaces versus tool and command interfaces}
\label{shared-tool-command-interfaces}

Affora is complementary to structured agent interfaces such as WebMCP, APIs, and command-line tools. These mechanisms can provide more direct and efficient machine action by avoiding visual grounding and GUI operation. Affora addresses a different question: whether task state and interaction meaning remain represented on a surface that both people and agents can inspect.

The two approaches can coexist. Agents may act through structured tools while consequential state is projected back into a shared interface. The same distinction applies to command-line interaction: commands and textual outputs are highly machine-friendly, but a persistent interface can better expose relationships, progress, outstanding work, and recovery options over time.

This suggests a broader principle: machine-efficient action need not require a machine-only representation of state.

\subsubsection{7.4 Beyond the DOM: desktop interfaces and software for human--agent collaboration}
\label{beyond-the-dom}

The semantic substrate studied here is implemented through web representations, but the distinction between visible and machine-readable interface structure is not web-specific. Windows UI Automation exposes control properties and supported actions \citeproc{ref-microsoft2025uia}{(Microsoft 2025)}. macOS accessibility APIs and Linux AT-SPI likewise expose structured interface information \citeproc{ref-apple2015accessibility}{(Apple 2015)} \citeproc{ref-gnomeatspi}{(The AT-SPI2 Maintainers n.d.)}. Electron can expose Chromium's accessibility tree through platform accessibility support \citeproc{ref-electronaccessibility}{(Electron Contributors n.d.)}. These mechanisms suggest possible implementations of Affora's semantic substrate beyond the DOM; transfer to them has not been evaluated here.

The same principles can therefore be asked of desktop software: whether visible controls correspond to machine-operable elements, whether state and available actions are recoverable, and whether task state survives transitions between views or windows. Testing Affora across native and Electron applications would determine how far these rules generalize beyond the web.

Software for human--agent collaboration extends this design problem to one or more people and agents working together through shared interfaces and persistent task state. Explicit state alone may be insufficient when participants act on shared work: the interface may also need to show who made a change, who is responsible for the next step, and whether actions conflict or can be reversed. Supporting this coordination is a possible extension of Affora's state and continuity rules, but has not been implemented or evaluated here.

\subsubsection{7.5 Limitations}
\label{limitations}

The present evidence is bounded in several ways. We evaluate small- and mid-tier models rather than frontier systems, and the strongest visual-design results come from components and relatively short tasks. The independent applications are concentrated in administrative and commerce interfaces, so broader ecological validity remains untested.

The work also evaluates machine readers rather than human users. Accessibility conformance and preservation of familiar interface structure do not establish usability, trust, or supervisability for people. The executable checks likewise cover only properties that can be decided from the rendered artifact; they are a conformance floor rather than a predictor of task success.

Finally, the extensions to desktop interfaces and software for human--agent collaboration discussed above remain hypotheses. The current implementation and evaluation are primarily web-based.

\subsubsection{7.6 Conclusion}
\label{conclusion}

Affora shows that designing for computer-use agents does not require collapsing software into a machine-oriented interface. Agent-relevant controls, choices, state, and outcomes can be made explicit while visual expression remains flexible.

More broadly, the work points toward software whose state remains legible regardless of who operates it. Agents may act through graphical interfaces, structured tools, command lines, or APIs; what can remain shared is the interface's representation of what is possible, what has happened, and what state the system is now in. Affora offers one design-system approach toward that shared human--agent surface.

\section*{Generative AI Use Disclosure}
Generative AI tools were used for language editing, restructuring, and
implementation support during manuscript and artifact preparation. AI model APIs
were also used to run the reported agent evaluations. The author verified all
text, figures, references, code, experimental procedures, and reported results,
and retains full responsibility for the work.

\clearpage
\appendix
\section*{Appendix}
\label{appendix}

\FloatBarrier

\subsection{Principle index}\label{principle-index}

The following reference table separates each rule from its associated check and evidence status. A measured effect applies to the tested condition; observed and proposed requirements are not presented as experimentally validated. Layout applies P5 and P9 across page composition rather than adding a separate principle family. In the check column, S1--S8 are component checks, FC1--FC8 flow checks, K1--K3 site checks, and FP1--FP5 flow patterns.

\begin{table}[htbp]
\centering
\small
\setlength{\tabcolsep}{4pt}
\begin{tabular}{@{}P{0.12\linewidth}P{0.36\linewidth}P{0.16\linewidth}P{0.29\linewidth}@{}}
\toprule
evidence & principle & scope & deciding check / implementation \\
\midrule
measured & P1 State is stated & component & S4, S7 \\
measured & P2 Presence over disclosure & component & S5 \\
measured & P3 Every affordance is named, in text & component & S1, S2 \\
measured & P7 Feedback persists & component & S7 · FP4 \\
measured, null & P8 One concept, one word & component + site & S2, K2 \\
measured & F1 Depth is the price & flow & FC1 · FP1 \\
measured & F2 Every screen states its own state & flow & FC2, FC3 · FP3 \\
measured / proposed & F4 Reversal over confirmation & flow & FC5 · FP5 \\
measured & F5 Dual paths: guided and direct & flow & FC6 · FP1 \\
measured & F6 Failures name their remedy & flow & FC7 · FP4 \\
measured & SC2 One glyph per function, name travels & site & K2 \\
observed & P4 Meaning never rests on colour or geometry alone & component & S8 \\
observed & P5 Visual hierarchy mirrors semantic hierarchy & component + layout & S3 \\
observed & P9 Structural isomorphism & component + layout & S3, S5 \\
observed & P10 Destructive actions are marked & component & FC5 · FP5 \\
null & P6 Constraints are stated before the attempt & component + flow & S3, FC4 · FP2 \\
null & F3 Constraints precede attempts & flow & FC4 · FP2 \\
proposed & F7 Completion is stated & flow & FC8 \\
proposed & SC1 Stable semantic colour roles & site & K1 \\
null with history & SC3 Navigation consistency & site & K3 \\
\bottomrule
\end{tabular}
\end{table}

\FloatBarrier

\end{document}